\documentclass[aps,showpacs,prl,twocolumn,superscriptaddress]{revtex4}
\usepackage{epsfig}
\usepackage{amsfonts}
\usepackage{amsmath}
\usepackage[utf8]{inputenc}
\usepackage[colorlinks,bookmarks=false,citecolor=blue,linkcolor=blue,urlcolor=blue]{hyperref}

\newcommand{\beq}{\begin{equation}}
\newcommand{\eeq}{\end{equation}}
\newcommand{\bea}{\begin{align}}
\newcommand{\eea}{\end{align}}
\newcommand{\nn}{\nonumber}

\newcommand{\dagga}{{\phantom{\dagger}}}

\begin{document}
\title{Renormalization group approach for the scattering off a single Rashba impurity in a helical liquid}
\author{Fran\c{c}ois Cr\'epin}
\affiliation{Institute for Theoretical Physics and Astrophysics,
University of W\"urzburg, 97074 W\"urzburg, Germany}
\author{Jan Carl Budich}
\affiliation{Institute for Theoretical Physics and Astrophysics,
University of W\"urzburg, 97074 W\"urzburg, Germany}
\author{Fabrizio Dolcini}
\affiliation{Dipartimento di Scienza Applicata e Tecnologia, Politecnico di Torino, 10129, Torino, Italy}
\author{Patrik Recher}
\affiliation{Institute for Mathematical Physics, TU Braunschweig, 38106 Braunschweig, Germany}
\author{Bj\"orn Trauzettel}
\affiliation{Institute for Theoretical Physics and Astrophysics,
University of W\"urzburg, 97074 W\"urzburg, Germany}

\date{\today}

\begin{abstract}
The occurrence of two-particle inelastic backscattering has been conjectured in helical edge states of topological insulators and is expected to alter transport. In this Letter, by using a renormalization group approach, we provide a microscopic derivation of this process,  in the presence of a time-reversal invariant Rashba impurity potential. Unlike previous approaches to the problem, we are able to prove that such an effect only occurs in the presence of electron-electron interactions. Furthermore, we find that the linear conductance as a function of temperature exhibits a crossover between two scaling behaviors: $T^{4K}$ for $K>1/2$ and $T^{8K-2}$ for $K<1/2$, with $K$ the Luttinger parameter.   
\end{abstract}


\maketitle

{\it Introduction.}---  Since the prediction of the quantum spin Hall phase \cite{Kane05, Kane05b} in HgTe quantum wells \cite{Zhang06b}, transport measurements on these compounds have shown evidence of a quantized edge conductance $G = 2e^2/h$, thereby paving the way for non-local dissipationless transport in semiconductors at zero external magnetic field \cite{Konig07, Konig08, Roth09}.  In the simplest case of quantum wells with inversion symmetry, transport occurs through two counter-propagating edge channels that carry opposite spin-1/2 quantum numbers. Such helical liquids form a new class of 1D quantum liquids in the sense that they are protected by time-reversal symmetry against single-particle elastic backscattering \cite{Kane05b, Moore06, Zhang06c}. However, deviations from the quantized conductance arise in various situations, involving either a breaking of time-reversal symmetry -- by a magnetic impurity for instance -- or the interplay between a time-reversal invariant (TRI) external potential and a source of inelastic scattering. Inelastic single-particle backscattering \cite{Budich12, Schmidt12} and two-particle backscattering \cite{Kane05b, Moore06, Zhang06c, Japaridze10} are two examples of the latter. In this Letter, we focus on two-particle backscattering off a TRI impurity and report new results regarding the temperature scaling of conductance corrections. Our purpose is to derive the Hamiltonian for such a process starting with a minimal model of an interacting helical liquid coupled to a TRI potential. In particular, we focus on a Rashba spin-orbit potential \cite{Japaridze10, Budich12}, which can originate from fluctuations of an electric field perpendicular to the 2D electron gas  \cite{Rothe10}, and acts as a TRI effective magnetic field that couples right and left movers. 
In the recent literature, inelastic two-particle backscattering off an impurity was mostly studied phenomenologically, by postulating the generic form of the Hamiltonian due to symmetry considerations -- namely TRI and Pauli principle \cite{Kane05b, Moore06, Zhang06c, Oreg12},
\beq \label{eq:H_2p}
H_{\textrm{2p}}^{\textrm{in}} =  \gamma_{\textrm{2p}}^{\textrm{in}} \left[(\partial_x \Psi^\dagger_+) \Psi^{\dagger}_+(\partial_x\Psi^{\dagga}_-)\Psi^{\dagga}_-\right](x_0) + \textrm{H.c.},
\eeq 
where $\pm$ designate right and left movers respectively. A straightforward scaling analysis \cite{Kane92a, Kane92b} would lead to a temperature dependence of $T^{8K-2}$ for conductance corrections, with $K$ the Luttinger parameter, implying a $T^6$ behavior in the limit of weak interactions, $K \simeq 1$. These studies, however, do not explain how two-particle backscattering is generated at the microscopic level.  To our knowledge, the only microscopic explanation proposed so far is the one by Str\"om {\it et al}. \cite{Japaridze10}, already based on Rashba spin-orbit coupling. 
Their analysis, however, leads to the unphysical conclusion that  these processes are present even in the limit of vanishing interactions. Indeed, without interactions, two-particle backscattering can always be factorized to two uncorrelated single-particle elastic backscattering processes and does not affect transport. A satisfactory explanation of the effect is therefore still lacking.

In this Letter, we use a renormalization group (RG) approach to show how two-particle inelastic backscattering is generated from Rashba spin-orbit coupling and Coulomb interactions.  Upon integrating the flow equations, we are able to show that the effect only occurs in the presence of electron-electron interactions. Furthermore, we find a $K$-dependent crossover behavior for the temperature scaling of the conductance corrections, namely 
\beq
\delta G/G_0 \sim 
\left\{
\begin{array}{ll}
(a_0 T/v)^{4K}    & \mbox{ if $K>1/2$},\;\\
(a_0 T/v)^{8K-2}     & \mbox{ if $1/4<K<1/2$},
\end{array}
\right. \label{crossover}
\eeq
where $a_0$ is the inverse bandwidth and $v$ the interaction-renormalized Fermi velocity. Our analysis demonstrates that, in the limit of weak interactions,  two-particle inelastic processes, with a scaling of $T^4$, are a more important source of scattering than usually anticipated from phenomenology.

{\it Model.}--- We study an interacting 1D helical liquid  in the presence of Rashba spin-orbit coupling. We set $\hbar = 1$ and the Hamiltonian of the system is the sum of three terms, $H = H_0 + H_{I} + H_R$, given by 
\begin{align} \label{eq:Hamiltonian}
&H_0 = \int dx \; \sum_{\eta=\pm} \Psi^\dagger_\eta(x) (-i \eta v_F\partial_x - E_F) \Psi^{\dagga}_\eta(x), \\
&H_{I} = \iint dx dx' \Psi^\dagger_+(x) \Psi^{\dagger}_-(x') g_2(x-x')\Psi^{\dagga}_-(x')\Psi^{\dagga}_+(x), \nn \\
&H_R = \int dx  \;\alpha(x)\left[ (\partial_x \Psi^\dagger_+)\Psi^{\dagga}_- -\Psi^\dagger_+(\partial_x \Psi^{\dagga}_-)\right](x) + \textrm{H.c.}. \nn
\end{align}
Here, $\Psi^{\dagger}_+(x)$ and $\Psi^{\dagger}_-(x)$ are creation operators for right and left moving electrons, respectively. Both species carry spin-$1/2$ opposite quantum numbers and hence transform as $\mathcal{T}\Psi^{\dagger}_{\pm}(x) \mathcal{T}^{-1} = \pm \Psi^{\dagger}_{\mp}(x)$ under time reversal. $H_0$ entails a strictly linear spectrum, with a finite bandwidth, the size of the bulk band gap. $v_F$ is the Fermi velocity, $E_F$ the Fermi energy. Without loss of generality, we consider in $H_I$ only interactions   between electrons moving in opposite directions, since chiral interactions -- so-called $g_4$ terms -- only renormalize the Fermi velocity.  Finally, $H_R$ describes a linear Rashba spin-orbit potential likely to stem from fluctuations of a transverse  electric field \cite{Japaridze10, Budich12}. We emphasize that $H_R$ is, in a helical liquid, the time-reversal invariant Hamiltonian with the lowest scaling dimension, able to couple right and left-movers. Nevertheless it has no effect on transport as long as elastic scattering is concerned \cite{Moore06}. In the following, we consider a point-like impurity, that is, $\alpha(x) = \alpha \delta(x)$. We show with a RG calculation how two-particle inelastic backscattering is generated. First, we carry it out on the fermion partition function before treating interactions exactly using bosonization.  

\begin{figure} 
\centering
\includegraphics[width=9cm,clip]{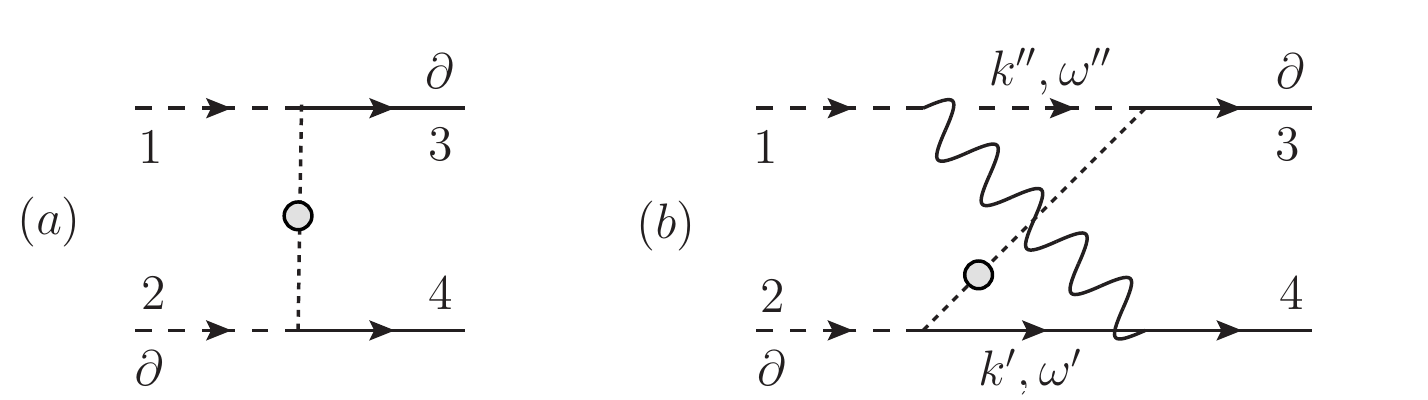}
\caption{Examples of diagrams to order $\alpha^2$ (a) and $g_2\alpha^2$ (b), in the expansion of the partition function. Partial derivatives signs indicate which external lines are differentiated with respect to $x$. Full (dashed) arrows are for right (left) movers. Wavy lines are for Coulomb interactions, and grey balls denote scattering off the impurity.}
\label{Fig:diagrams_2}
\end{figure}

{\it RG for interacting fermions.}--- Much insight is gained by first treating, at the fermion level, both interactions and the Rashba potential as perturbations to the non-interacting fixed point. We use the path integral representation of the partition function, $\mathcal{Z} = \int D\Psi_\pm^* D\Psi_\pm e^{-S}$, with an action $S = \int_0^\beta d\tau \left[\sum_{\eta=\pm} \int dx\Psi_\eta^*(x,\tau) \partial_\tau \Psi_\eta(x,\tau) + H(\tau) \right]$, and $\Psi_\pm(x,\tau)$, $\Psi_\pm^*(x,\tau)$ Grassmann fields. 
 We introduce an ultra-violet (UV) cutoff $\Lambda$ of the order of half the bulk band gap on the dispersion relation of both right and left movers as $v_F|\eta k - k_F| < \Lambda$, with $k_F$ the Fermi momentum. Following Ref. \onlinecite{Shankar94}, we then proceed to integrate out the fields living on an infinitesimal momentum shell $\Lambda/s < v_F|\eta k - k_F| < \Lambda$, with $s = 1 + d\ell$. As usual in a 1D quantum liquid, interactions contribute an infinite series of diagrams. However, in the absence of $2k_F$ scattering processes, and the impurity being point-like, $g_2$ is invariant under RG transformations. The integration of high energy fields also generates new terms. To third order perturbation theory, the diagram (b) depicted in Fig. \ref{Fig:diagrams_2} generates an inelastic two-particle backscattering process whose action is of the form
\begin{align}
&S_{\textrm{2p}} = \gamma_{\textrm{2p}} \int \prod_{i=1}^4\frac{dk_i}{2\pi}\frac{d\omega_i}{2\pi} 2\pi \delta(\omega_1+\omega_2-\omega_3-\omega_4) \times \nn \\
&(-ik_3\Psi^*_+(3))\Psi^*_+(4) (ik_2\Psi_-(2)) \Psi_-(1) + \{+ \leftrightarrow - \}. \label{eq:S_2p}
\end{align}
This is precisely the action one would derive from the Hamiltonian of Eq. \eqref{eq:H_2p}, in momentum space. The scaling dimension of $\gamma_{\textrm{2p}}$ is -3 by power-counting, and, taking into account the aforementioned third-order diagram, its flow equation is
\beq
\label{eq:RG_flow_1}
\frac{d\gamma_{\textrm{2p}}}{d\ell} = -3 \gamma_{\textrm{2p}}(\ell) + \frac{\alpha(\ell)^2}{v_F \Lambda} \frac{g_2}{2\pi v_F}.
\eeq
Note that the initial condition is $ \gamma_{\textrm{2p}}(\ell=0)=0$, since two-particle inelastic backscattering is absent from the bare action. Finally, power counting on the Rashba action yields for $\alpha(\ell)$ the flow equation
\beq
\label{eq:RG_flow_2}
\frac{d\alpha}{d\ell} = -\alpha(\ell).
\eeq
This calculation confirms that inelastic two-particle backscattering from a Rashba impurity is only generated in the presence of Coulomb interactions, as it disappears altogether as soon as $g_2 = 0$. We emphasize that to second order in $\alpha$, diagrams such as (a) in Fig. \ref{Fig:diagrams_2} do not generate inelastic processes since Matsubara frequencies are conserved independently at each Rashba scattering vertex; in this example, $\omega_1 = \omega_3$ and $\omega_2 = \omega_4$. Finally, we point out that $g_4$ (chiral) interactions fail to generate inelastic two-particle backscattering as all diagrams will be suppressed by the Pauli principle.  

{\it Bosonization.}--- We refine our analysis by treating interactions exactly, through bosonization of the fermion Hamiltonian.  Excitations around the true ground state of the 1D interacting helical liquid are indeed described by the Luttinger liquid Hamiltonian, $\mathcal{H}_0 = H_0 + H_I$, with
\beq
\mathcal{H}_0 = \frac{v}{2\pi} \int dx \; \left[ K (\partial_x \theta)^2 +  \frac{1}{K}(\partial_x \phi)^2 \right],
\eeq
where $\phi$ and $\theta$ are two boson fields describing density and quantum phase fluctuations, respectively \cite{Haldane81}, and obeying the following commutation relation $[\phi(x),\partial_x\theta(x')]=i\pi \delta(x-x')$. For repulsive electrons, $K<1$, while $K=1$ in the non-interacting case. By using the bosonization identity $\Psi_{\pm}(x)= \kappa_\pm (2 \pi a)^{-1/2} e^{\pm i k_F x} e^{-i(\pm \phi-\theta)}$, the bosonized form of the Rashba Hamiltonian is readily obtained as \cite{Japaridze10, Budich12} 
\begin{align}
\label{eq:HR}
&H_R = i\kappa_+ \kappa_- \int dx \frac{\alpha(x)}{\pi a} \left( \frac{2\pi a}{L}\right)^K  \times \nn \\
&\times :\partial_x\theta(x) \left( :e^{-i2\phi(x)}: e^{i2k_Fx} + :e^{i2\phi(x)}: e^{-i2k_Fx} \right):
\end{align}
where $:\ldots:$ indicate normal order with respect to boson operators that annihilate the ground state of the helical liquid. Here, $\kappa_\pm$ are Klein factors \footnote{We use the Majorana fermions representation.}, and $a$ is a short-distance cutoff and the running scale in the RG approach.  
For all purposes here, its bare value $a_0$ can readily be  identified with $\Lambda^{-1}$, where $\Lambda$ is the bandwidth previously introduced in the fermion RG analysis. The total bosonized Hamiltonian of the system is $H = \mathcal{H}_0 + H_R$. We perform an RG transformation in real space \cite{CardyBook}, which consists in rescaling first the short distance cutoff, $a \rightarrow a' = (1+d\ell)a$, and then the couplings in order to keep the low-energy form of the Hamiltonian invariant. We rescale the cutoff order by order in an expansion to order $\mathcal{O}(\alpha^2)$ of the partition function $\displaystyle \mathcal{Z} = \textrm{Tr} e^{-\beta\mathcal{H}_0 } \hat{U}(\beta,0)$, where $\hat{U}(\beta,0) = \mathcal{T}_\tau e^{- \int_0^\beta d\tau_1 \hat{\mathcal{H}}_R(\tau_1) }$ is the time-evolution operator in the interaction representation. 
At tree level, we derive the following flow equation for the Rashba coupling
\beq
\label{eq:rg_alpha}
\frac{d\tilde{\alpha}}{d\ell} = - K \tilde{\alpha}(\ell),
\eeq
in which we have introduced the dimensionless variable $\tilde{\alpha} = \alpha/(\pi v a)$. Bosonization readily takes into account vertex corrections due to interactions and we recover Eq. \eqref{eq:RG_flow_2} in the limit of weak interactions, $K \rightarrow 1$. Two-particle inelastic backscattering is generated as a second-order perturbation process. Indeed the expansion to order $\alpha^2$ of the partition function leads to a term
\begin{align}
\label{eq:fermions}
&\alpha^2 \left(\frac{2a}{v}\right) \int_0^\beta d\tau_1 \; \left[(\partial_x \hat{\psi}^\dagger_{+})\hat{\psi}^\dagger_{+} (\partial_x\hat{\psi}_{-}) \hat{\psi}_{-}\right](1) \nn \\
&+\frac{\alpha^2}{2}\int_{v|\tau_1 - \tau_2|>a} \hspace{-1.1cm}d\tau_1 d\tau_2\;  (\partial_x \hat{\psi}^\dagger_{+})(1)\hat{\psi}^\dagger_{+}(2) (\partial_x\hat{\psi}_{-})(1)\hat{\psi}_{-}(2) + \textrm{H.c.}
\end{align}
where the UV cutoff is enforced by splitting the double integral over imaginary time into two parts for which $v|\tau_1 - \tau_2|<a$ and $v|\tau_1 - \tau_2|>a$, respectively. The first line, corresponding to short time differences $\tau_1 \simeq \tau_2$, contributes an inelastic scattering process. Importantly, in the limit of vanishing interactions, the first term exactly cancels a similar term generated by the cutoff rescaling in the second integral, proving that no two-particle backscattering occurs without interactions~\footnote{This subtlety of the RG procedure for impurity scattering was first noticed by Giamarchi and Schulz in their study of Anderson localization in 1D interacting liquids~\cite{Giamarchi88}, and recently emphasized by Gornyi {\it et al.}~\cite{Polyakov07} in their treatment of weak localization. }. By writing  Eq.~\eqref{eq:fermions}  in terms of the bosonic fields and after normal ordering, we obtain 
\begin{align}
\label{eq:OPE}
&:\partial_x\theta(1)  :e^{i2\sqrt{K}\phi(1)}::\times :\partial_x\theta(2)  :e^{i2\sqrt{K}\phi(2)}:: \; \; =\nn \\
&\frac{1}{2} \left(\frac{2\pi}{L}(y+ a) \right)^{2K} \frac{1-2K}{(y + a)^2}  :e^{i2\sqrt{K}\phi(1)} e^{i2\sqrt{K}\phi(2)}: + \ldots
\end{align}
Note that we have rescaled the bosonic fields according to $\sqrt{K}\theta \rightarrow \theta$ and $\phi/\sqrt{K} \rightarrow \phi$. Furthermore, $y_{1(2)} = v\tau_{1(2)}$ has dimension of a length and we define $y = y_1-y_2$. Dots represent extra terms that have a vanishing expectation value. Keeping the lowest order term in an operator product expansion, the rescaling of $a$ generates a new coupling, which we identify with a two-particle inelastic backscattering process. At the end of the RG step, the time-evolution operator $\hat{U}(\beta,0)$ is corrected by a Hamiltonian 
\begin{align}
\hspace{-0.2cm}\int_0^\beta d\tau_1 \hat{\mathcal{H}}_{\textrm{2p}}(\tau_1) =\frac{\tilde{\gamma}_{\textrm{2p}}}{a}  \int_0^{v\beta} dy\left[ e^{i4\sqrt{K}\phi(x_0,y)} + \textrm{H.c.} \right]
\end{align}
where $\tilde{\gamma}_{\textrm{2p}}$ is a dimensionless coupling~\footnote{Compared to Eq. \eqref{eq:H_2p}, we have $\tilde{\gamma}_{\textrm{2p}}^{\textrm{in}}=\gamma_{\textrm{2p}}^{\textrm{in}}/(\pi^2 v a^2) $  } given by
\beq
\label{eq:true_gamma}
\tilde{\gamma}_{\textrm{2p}}(\ell) = \tilde{\gamma}^{\textrm{in}}_{\textrm{2p}}(\ell) - \frac{\tilde{\alpha}(\ell)^2}{2K}(1-2K).
\eeq
On the r.h.s of Eq. \eqref{eq:true_gamma}, $ \tilde{\gamma}^{\textrm{in}}_{\textrm{2p}}(\ell)$ stands for the true inelastic backscattering processes, which in the present case, has the bare value $ \tilde{\gamma}^{\textrm{in}}_{\textrm{2p}}(\ell=0)=0$. The second term is the correction arising from the first integral in Eq. \eqref{eq:fermions}. Using Eq. \eqref{eq:rg_alpha}, the flow equation for the true inelastic two-particle backscattering reads
\beq
\label{eq:rg_gamma}
\frac{d\tilde{\gamma}^{\textrm{in}}_{\textrm{2p}}}{d\ell}  = (1-4K) \tilde{\gamma}^{\textrm{in}}_{\textrm{2p}}(\ell) + \left(1-\frac{1}{K} \right)(1-2K)\tilde{\alpha}(\ell)^2 .
\eeq
In the absence of interactions, $K=1$ and $ \tilde{\gamma}^{\textrm{in}}_{\textrm{2p}}(\ell)=0$ at any scale $\ell$, consistently with the fermionic result of Eq. \eqref{eq:RG_flow_1}.

\begin{figure} 
\includegraphics[width=8cm,clip]{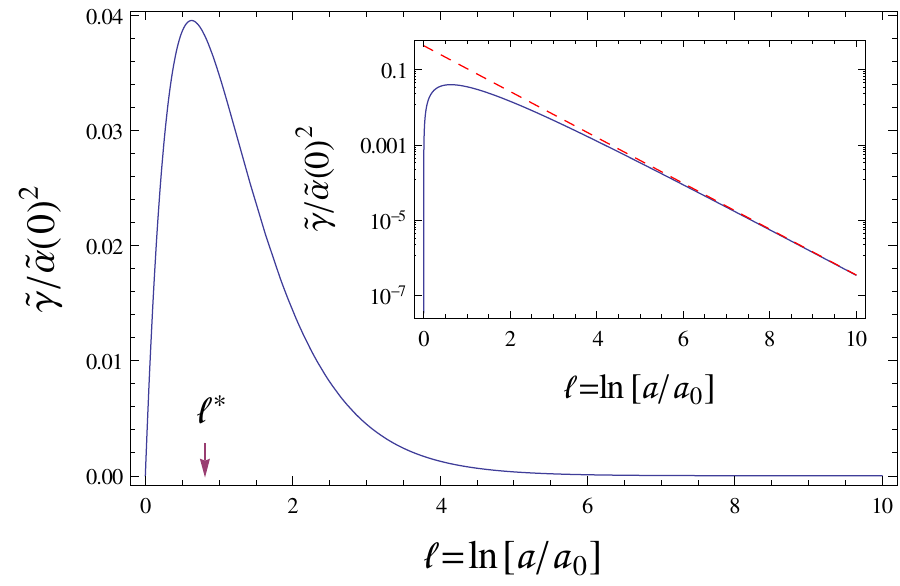}
\caption{Flow of $\tilde{\gamma}^{\textrm{in}}_{\textrm{2p}}$ as a function of $\ell = \ln [a/a_0]$, for $K=0.7$. A crossover scale $\ell^*$ separates a region of linear growth for small $\ell < \ell^*$ from a region of exponential decay at large $\ell > \ell^*$. The inset is the same plot on a semi-logarithmic scale. The asymptote is for $e^{-2K\ell}$.  }
\label{Fig:flow1}
\end{figure}
{\it Transport.}---   We now apply a  small voltage bias $V$ to the helical liquid. The dc conductance $G$ is then obtained from linear response as the zero-frequency limit of the current-current correlation function \cite{Kane92a, Kane92b}. The latter is evaluated in perturbation theory, and because of TRI, corrections to the quantized conductance $G_0 = e^2/h$  arise only to order $\mathcal{O}(\alpha^4)$. Equivalently here, by letting the system flow to a certain scale $\ell$ we obtain corrections to conductance to order $\mathcal{O}(\tilde{\gamma}^{\textrm{in}}_{\textrm{2p}}(\ell)^ 2)$ in perturbation theory. Integrating equations \eqref{eq:rg_alpha} and \eqref{eq:rg_gamma} between $0$ and $\ell$ we find $\tilde{\alpha}(\ell) = \tilde{\alpha}(0) e^{-K \ell}$ and
\beq
\tilde{\gamma}^{\textrm{in}}_{\textrm{2p}}(\ell) = \left(1-\frac{1}{K} \right)\tilde{\alpha}(0)^2 \left[ e^{(1-4K)\ell} - e^{-2K\ell} \right],
\eeq
which we plot in Fig.~\ref{Fig:flow1} for a particular value of $K$.   We find two different asymptotic behaviors, separated by a crossover scale $\ell^* = (2K-1)^{-1}\ln[(4K-1)/(2K)]$, independent of $\tilde{\alpha}(0)$, the bare value of the Rashba coupling. For small $\ell \ll \ell^*$, $\tilde{\gamma}^{\textrm{in}}_{\textrm{2p}}(\ell) \simeq \tilde{\alpha}(0)^2 (1-K^{-1})(1-2K)\ell e^{-2K\ell} $  while for $\ell \gg \ell^*$,
$\tilde{\gamma}^{\textrm{in}}_{\textrm{2p}}$ crosses over from $\tilde{\gamma}^{\textrm{in}}_{\textrm{2p}}(\ell) \simeq -\tilde{\alpha}(0)^2 (1-K^{-1})e^{-2K\ell}$ for $K>1/2$ to $\tilde{\gamma}^{\textrm{in}}_{\textrm{2p}}(\ell) \simeq \tilde{\alpha}(0)^2 (1-K^{-1})e^{(1-4K)\ell}$ for $K<1/2$. As can be seen from Eq. \eqref{eq:rg_gamma}, $K=1/2$ is an intermediate fixed point where two-particle inelastic backscattering is not generated, at least not in second order perturbation theory. Integrating out energy scales between the bare cutoff $a_0$ and the thermal length $a(\ell) = v\beta$, we obtain the   temperature scaling of Eq. (\ref{crossover}) for conductance corrections to order $\mathcal{O}(\tilde{\alpha}^4)$,
 for temperatures lower than the crossover temperature $T^* = (v/a_0)e^{-\ell^*}$, while for $T> T^*$, these corrections are logarithmically suppressed as $T$ approaches $T_0 = v/a_0$
\beq
\delta G/G_0 \sim (a_0 T/v)^{4K} \ln^2 (a_0 T/v),
\eeq 
for all values of $K$. Note that for $K<1/4$, two-particle backscattering becomes a relevant perturbation and the Rashba impurity effectively cuts the helical liquid into two separate regions \cite{Kane92a, Kane92b, Zhang06c}. We emphasize that at low temperatures, $T\ll T^*$, and in the limit of weak interactions, $K \simeq 1$, we predict that corrections to the conductance from two-particle backscattering off a Rashba impurity scale as $T^4$ instead of $T^6$, as one would naively predict from the Hamiltonian of Eq. \eqref{eq:H_2p}.

{\it Discussion.}---  It is worth emphasizing the difference with respect to a recent work by Schmidt {\it et al}.~\cite{Schmidt12}, where a different model for a helical liquid  with broken $S_z$ symmetry was analysed. There, Rashba spin-orbit coupling, by imposing a momentum-dependent rotation of the spin of right and left movers, allows for inelasic {\it single-particle} backscattering off a scalar impurity. These processes contribute a $T^4$ correction to the quantized conductance, in the limit of weak Coulomb repulsion. The fact that in our approach, {\it  two-particle} backscattering actually leads to the same temperature dependence is a mere coincidence. 


{\it Conclusions.}--- In summary, we have studied the simplest model of a 1D helical liquid in the presence of a TRI impurity and electron-electron interactions, that alters transport. Our approach provides a firm microscopic explanation for the generation of two-particle backscattering in helical liquids 
and predicts the occurrence of a conductance crossover, which could not be captured by previous approaches. As current estimates for the Luttinger parameter in HgTe quantum wells, ranging between $K\simeq0.5$ and $K\simeq1$, show a strong dependence on the geometry of the device \cite{Maciejko09, Kane09, Chamon09, Japaridze10}, all regimes presented here could be of experimental relevance in transport measurements.\\
 
{\it Acknowledgements.}--- We thank Pauli Virtanen for many interesting discussions. Furthermore, we acknowledge financial support from the German-Italian Vigoni Program as well as the DFG.

\bibliography{Top_ins_wurzburg}

\end{document}